\documentclass[12pt]{article}

\usepackage{scicite}

\usepackage{times}
\usepackage{color}
\usepackage{wasysym}
\usepackage{amssymb}

\usepackage{graphicx}

\newenvironment{sciabstract}{%
\begin{quote} \bf}
{\end{quote}}

\newcommand{\be}{\begin{equation}}
\newcommand{\ee}{\end{equation}}

\newcounter{lastnote}

\title{Universal Signatures of Fractionalized Quantum Critical Points}

\author
{Sergei V. Isakov,$^{1}$  Roger G. Melko,$^{2}$ Matthew B. Hastings$^{3 \ast}$\\
\\
\normalsize{$^{1}$Theoretische Physik, ETH Zurich, 8093 Zurich, Switzerland,}\\
\normalsize{$^{2}$Department of Physics and Astronomy, University of Waterloo, Ontario, N2L 3G1, Canada}\\
\normalsize{$^{3}$Duke University, Department of Physics, Durham, North Carolina, 27708,}\\
\normalsize{Microsoft Research, Station Q, CNSI Building, University of California, Santa Barbara, California, 93106}\\
\\
\normalsize{$^\ast$To whom correspondence should be addressed; E-mail: xhastings@gmail.com.}
}

\date{}

\begin{document} 

\baselineskip24pt

\maketitle

\begin{sciabstract}
Groundstates of certain materials can support exotic excitations 
with a charge that's a fraction of the fundamental electron charge.  
The condensation of
these fractionalized particles has been predicted to drive novel quantum phase
transitions, which haven't yet been observed in realistic systems. Through numerical 
and theoretical analysis of a physical model of interacting lattice bosons, we
establish the existence of such an exotic critical point, called XY*.  We
measure a highly non-classical critical exponent $\eta = 1.49(2)$, and construct a 
universal scaling function of winding number distributions that directly demonstrates 
the distinct topological sectors of an emergent $Z_2$ gauge field.  The universal quantities 
used to establish this exotic transition can be used to detect other fractionalized 
quantum critical points in future model and material systems.
\end{sciabstract}

\pagebreak

It is a remarkable fact that, in this age of high-energy accelerator experiments, certain types of 
fundamental quantum particles can only be studied in what seems the relatively pedestrian world of tabletop condensed-matter physics experiments.
Consider for example the familiar electron, carrying fundamental charge $e$.  Unlike the proton, whose charge originates from quarks with fractional charge, no energy is sufficiently high to break up the charge of an electron.  However, as demonstrated years ago by the measurement of fractional Hall conductance, if one places an electron in certain clean two-dimensional materials in a strong magnetic field, its charge can indeed break into fractions -- $e/3,e/5$, and so on\cite{Laughlin} -- each fractional charge arising from a {\it quasiparticle} emerging in the sample.

Such quasiparticles share all the important characteristics of real particles.  In ``deconfined'' phases of matter with a gap to excitations, the quasiparticles can be separated a large distance from each other, making them well-defined localized objects with a sharp energy-momentum dispersion relation.
Many types of quasiparticles are known to occur as elementary excitations in the low-temperature groundstates of quantum materials, most typically as the more mundane examples of normal bosons or fermions. \footnote{Although {\it anyon} excitations realizing more exotic braid statistics are also possible.}
The condition for their emergence is the realization of a particular exotic ``vacuum'', which is the groundstate of some condensed matter system. It is as excitations out of this strange vacuum that these unfamiliar quasiparticles can flit in and out of existence.  

For over a decade, condensed matter physicists have searched for delicate and elusive fractionalized particles in systems other than Hall effect materials.  Theoretical predictions have identified a class of low-temperature paramagnets, the quantum spin liquids, as holding particular promise for supporting them \cite{RK1,SL1,Tang}.
However, experimental searches for these fractional charges and their parent spin liquid vacuum remain unconvincing.  This is due in part to the difficulty in constructing measurements that are able to identify the experimental signatures necessary to betray their existence in the variety of specific materials -- ranging from fabricated solid state devices, to delicate organic magnets -- thought to harbor possible spin liquid states\cite{LeonSL}.  
Perhaps the most interesting recent experimental candidate is a set of materials that may display fractional particles with a gapless Fermi surface\cite{Yamashita,dwavebose,organic1,organic2}.  This leads to an even more tenuous situation: fractionalized quasiparticles {\it without} an excitation gap, where interaction between quasiparticles makes it problematic even to define a fractionalized excitation. \footnote{However, in the case of Refs.~\cite{Yamashita,dwavebose,organic1,organic2}, Fermi liquid theory suggests that fractionalized excitations remain sharp near the surface.} 
In bosonic systems, such interacting gapless fractional quasiparticles have been proposed to mediate 
novel quantum critical points that exist in certain order-to-order transitions \cite{senthilscience}.  
These ``Landau-violating'' critical points (which rely on long-wavelength fluctuations of fractional particles)
have the advantage that signatures of the fractionalization are manifest in {\it universal} quantities, such as critical exponents, avoiding reliance on measurements of specific material-dependent quantities.

A major goal of the theoretical community has been to demonstrate the existence of these ``deconfined'' quantum critical points in realistic microscopic models, a task that necessarily falls to large-scale numerical simulation --  quantum Monte Carlo (QMC).  A variety of models which may contain such order-to-order critical points (the best candidate being Sandvik's J-Q model \cite{JQ})
are currently under intense numerical study. 
In this paper, we study deconfined quantum criticality in a different context -- an order-to-{\it disorder} transition between a superfluid and a gapped spin liquid state, in a physical model of lattice bosons.
We establish the existence of fractionalization at this critical point by confirming the prediction \cite{XYstar1,XYstar2,grover, SergeiEta}
of a strongly non-classical universal critical exponent $\eta$. 
As suggested in Ref.~\cite{senthilscience}, the defining property of interacting fractionalized excitations  
is the presence of an emergent gauge field.  
By considering {\it topological} properties of this gauge field, which enforce nonlocal constraints, 
we construct the first direct measurement of fractionalized excitations at a critical point. 
These topological properties manifest themselves in new universal scaling functions that can be used as a test of the existence of fractionalized excitations in a variety of other models and materials.

\begin{figure}
\centerline{\includegraphics[width=0.85\columnwidth]{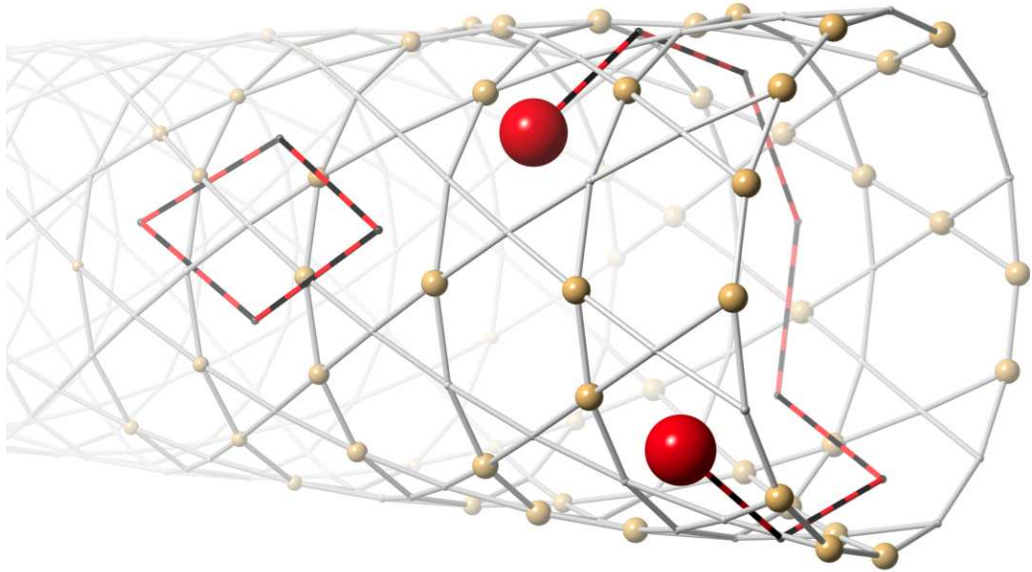}}
\caption{ A section of the toroidal kagome-lattice simulation cell.  Gold spheres label sites occupied by bosons in a representative configuration.  Red spheres are fractional charges, marking ``defect hexagons" (those which do not have three bosons per site).  Defect hexagons are shown joined by an open string.  We also show a representative closed string not associated with fractionalized particles.
}
\label{fig:lattice}
\end{figure}

{\bf The Kagome Bose-Hubbard Model.}  We examine a variant of the Bose-Hubbard Hamiltonian, introduced in Ref.~\cite{IHM}, which is a simple model of hard-core bosons hopping (with strength $t$) on a two-dimensional kagome lattice with an energetic constraint ($V$) favoring three bosons per hexagonal lattice plaquette:
$
H = -t \sum_{\langle ij \rangle} [b^{\dagger}_i b_j + b_i b^{\dagger}_j] + V \sum_{\hexagon} (n_{\hexagon})^2. 
$
This model has been convincingly shown to contain a superfluid to spin-liquid quantum phase transition at $V/t \approx 7$
through a smoking-gun measurement of the topological entanglement entropy\cite{IHM}.  
In that work, two universal exponents were measured at the critical point, related to the divergence of the correlation length ($\nu$) and the isotropy of the space and time dimensions ($z$), both of which fall in the conventional 3D XY universality class.  
A section of the toroidal kagome lattice geometry used in our QMC simulations, along with a representative boson configuration,
is shown in Fig.~\ref{fig:lattice}.

This system conserves total {\it charge} (the boson particle number).
Deep in the spin liquid phase  ($V\gg t$), it is known that exotic quasi-particles carrying half-odd integer charge exist as  fundamental excitations out of the groundstate vacuum.
At infinite $V$ and at half filling, the system is restricted to a space of states with three particles in each hexagon.  In this limit the model can be mapped to a dimer model on a triangular lattice formed by the centers of the hexagons, with three dimers per site\cite{BFG}.   
Turning on a non-vanishing $t/V$ produces virtual excitations out of this space of states, leading to effective exchange terms which give a quantum spin liquid phase.  Imagine removing a single particle from this state, creating a pair of defect sites in the dimer model with only two dimers.  In a phase with mobile, deconfined defects, these defects can become separated by large distances, and since the state has total charge $-1$ relative to the vacuum, each defect carries charge $-1/2$.  Similarly, defects with four dimers carry charge $+1/2$.  The model is thus in a $Z_2$ spin liquid phase, but electric defects carry half-odd charge.\footnote{This can be viewed as a topological quantum field theory, with particles $1,e^{+1/2},e^{-1/2},m,...$, where $e^{\pm 1/2}$ denotes the charge on the electric particle and the $...$ denote an infinite sequence of additional particles with charge shifted by any integer.}  Overlaying this configuration with a fixed reference configuration gives a transition graph, which is a gas of strings, with open strings connecting defects\cite{rk}.   There are three dimers per site so strings may intersect; in Fig.~\ref{fig:lattice} we only show two strings for clarity.  Since links in the string arise alternately from the given configuration and the reference configuration, links need not have a particle on them in the image shown.

Colloquially, one can view the fractionalized particles as a ``square-root'' of the boson field, with the $Z_2$ gauge field present due to the sign
ambiguity in the square-root.  Let  $b^\dagger,b$ denote creation and annihilation operators for the real physical bosons, and let $\phi^\dagger,\phi$  denote the same operators for the fractionalized particles.  At an $XY^*$ critical point, the fractionalized field $\phi$ undergoes an ordinary $XY$ transition.
As pointed out in Refs.~\cite{grover,earlyXYstar}, this dramatically affects the observed critical exponents, since the order parameter $b$ usually associated with the transition is actually a {\it composite} operator, made up of two $\phi$ fields.  This leads to the same exponent $\nu$ as in the ordinary $XY$ critical point.  However, the exponent $\eta$ controlling the equal time ground state correlation function $\langle b^\dagger(0) b(x) \rangle$ is significantly modified, leading to
$\eta\approx  1.45$ as estimated by field theoretic and Monte Carlo simulations of the correlation of a composite operator in the 3D XY model\cite{compositefieldtheory,compositeMC}.
This contrasts strikingly with $\eta\approx 0.03$ in the ordinary 3D XY transition.

\begin{figure}
\centerline{\includegraphics[width=0.64\columnwidth]{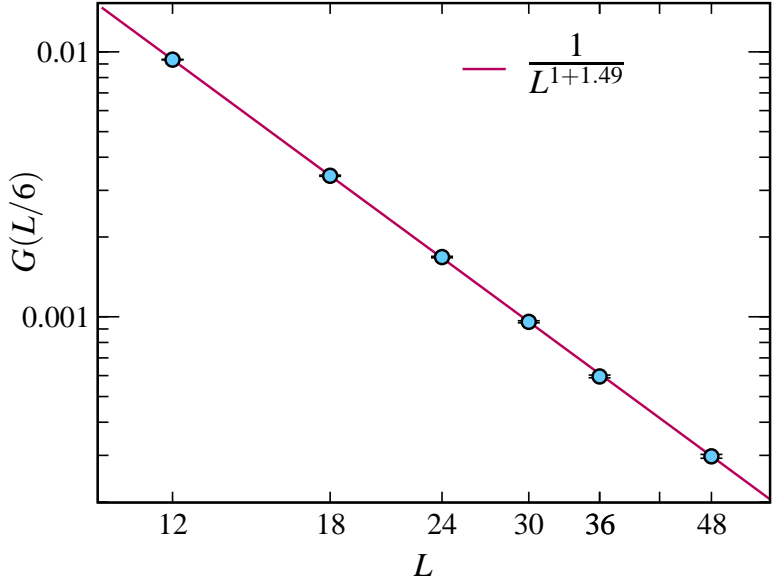}}
\caption{ 
Equal time Green's function $G(L/6) = \langle b^\dagger(0,0) b(L/6,0) \rangle$ as a function of system size $L$.
}
\label{fig:eta}
\end{figure}

The measurement of $\eta$ in the kagome model is challenging, since QMC does not directly access the superfluid order parameter operators necessary to measure the relevant correlation function.  To overcome this fact, we have measured the equal time Green's function in real space: $G(x)\equiv G(\tau=0,r) = \langle b^\dagger(0) b(r) \rangle$ \cite{WormA,gfsse}.  This measurement
involves keeping track of the defects created in the non-local {\it loop} algorithm\cite{SSE} as it traverses the QMC's $d+1$ space-time simulation cell.  At the critical point, $G(r)$ should decay at sufficiently long distances as $1/r^{1+\eta}$.  Finite size effects can be minimized by looking at $G(L/m)$, which decays as $1/L^{1+\eta}$, where $L$ is the linear system size and $m$ is some (fixed) number. In Fig.~\ref{fig:eta}, we show $G(L/6)$ as a function of $L$.  The Green's function decays algebraically with $\eta=1.49(2)$, a value that is consistent with $\eta$ for a composite operator in the 3D XY model.  This strongly non-classical $\eta$ shows that the boson operator $b$ is indeed a composite operator of fractionalized excitations.

\begin{figure}
\centerline{\includegraphics[width=0.7\columnwidth]{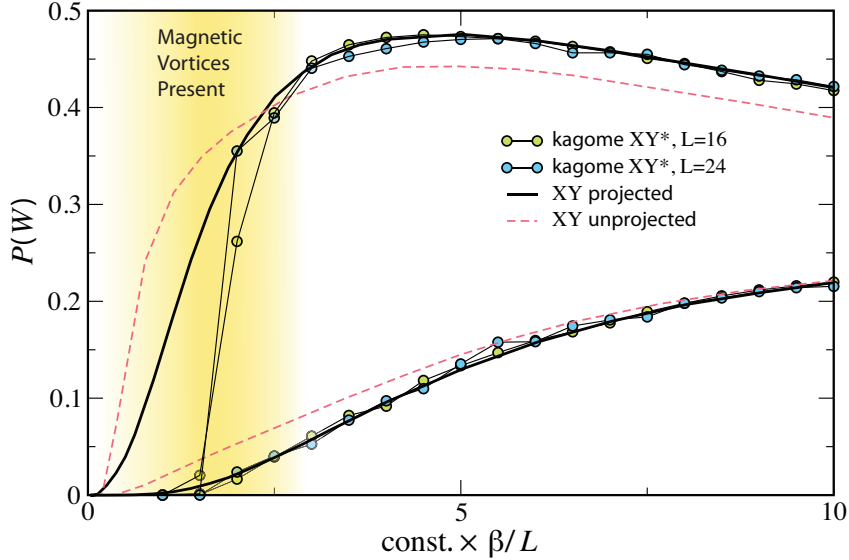}}
\caption{ Comparison of winding number distribution $P(W_x)$ for the kagome-lattice XY* transition at $L=16,24$, to $P_{even}(2W_x)$ for a conventional $L=16$ square-lattice $XY$ transition (described in the Supporting Online Material, Section 1.1).
Dashed lines show the unprojected distribution of $XY$ winding numbers $P(W)$ for comparison.  Upper curves are $W=1$ and lower are $W=2$.
The $x$-axis is $\beta/L$ for the kagome lattice, and $\beta/L$ times a non-universal velocity ratio for the $XY$ model.
}
\label{fig:wind}
\end{figure}

More direct evidence  can be obtained from topological properties, as we now show by constructing new scaling functions that exploit universality, and as such can be used as a test for the existence of fractionalized excitations in other models and experiments.   
Consider a path in imaginary time in which a total of $n_x$ of the $\phi$ particles wind around the $x$-direction of the torus geometry.  In this case, the total charge $W_x$ that winds around the torus is equal only to $n_x/2$.  However, the only closed paths in imaginary time that are allowed are those in which $n_x$ is even, as the physical charge $W_x$ winding around the torus must be an integer.  This is a manifestation of topological properties of the $Z_2$ gauge field.  The gauge field is gapped, so does not affect the dynamics of the $\phi$ field, but there are a total of eight different topological sectors (two sectors in each space direction and one in the time direction) and summing over sectors restricts to even winding and even number of the fractionalized particles (see Fig.~\ref{fig:lattice}, where winding a particle drags a string, changing the topological sector).   It is known that the winding number variance $\langle W_x^2 \rangle$ in the ordinary $XY$ model is a universal function of $L/v \beta$, where $L$ is the length scale of the torus and $v$ is a non-universal velocity.  It is natural (and we have confirmed numerically) that the full winding number distribution, as quantified by the probabilities $P(W_x)$ of observing a given winding, is also a universal function.
Thus, the universal critical properties of the winding number distribution $W_x$ in our model can be computed from a model undergoing an ordinary $XY$ transition in $2+1$ dimensions, by restricting to sectors with even winding numbers $n_x,n_y$ and even total particle number.
Then, one can determine the probability of observing given $n_x$ with
\be
P(W_x)=P_{even}(2W_x),
\ee
where $P_{even}(2W_x)$ is the probability, after projecting onto these even sectors, of observing $n_x=2W_x$.
While one does not know the scaling functions controlling $P_{even}$ analytically, we have sampled them using quantum Monte Carlo on a different model which contains a conventional XY transition (Supporting Online Material, Section 1.1).   As shown in Fig.~\ref{fig:wind}, the prediction of the theory shows excellent agreement for two distinct scaling functions (we also tested $P(3)$ but it is not shown) over the majority of the observed curve.  The region in which the functions do not agree, at higher temperatures, is due to low energy magnetic excitations (Supporting Online Material, Section 1.2); one can observe that as system size is increased, the range of agreement improves.

This provides a direct test of the fractionalized charge.  The fit of all the scaling functions uses only a single adjustable parameter, the non-universal velocity ratio $v_{XY}/v_{kagome}$.  In the limit of $v \beta/L\rightarrow \infty$, the winding numbers become large, and both the projected and unprojected winding number distributions converge to approximate Gaussians, so the projection onto even numbers has little effect on the variance.  Then, one will observe $\langle W_x^2 \rangle_{XY^*}= (1/4) \langle W_x^2 \rangle_{XY}$, where the subscripts denote the two different universality classes and we assume both models are at the same ratio $v\beta/L$.  However, testing this agreement is difficult as it requires an accurate measurement of the non-universal velocity.  In fact, the best way we have found to measure the non-universal velocity (as the high temperature measurements are complicated by the magnetic excitations) is to study the full distribution of scaling functions which provides a much more stringent test.
Note also that the scaling functions of the unprojected $XY$ model are very different from the projected functions at moderate $v/L\beta$, where both functions are strongly non-Gaussian and non-Poissonian.

{\bf The Phase Diagram.} As mentioned above, Fig.~\ref{fig:wind} shows slow convergence to universal scaling in the high temperature critical regime (small $v\beta/L$).  To understand this, we have considered the phase diagram of the model as a function of $V/t$ and temperature $T$.  At small $V/t$, the model is superfluid at zero temperature,
with a second order transition at non-zero temperature.  At larger $V/t$, but still below the critical point, the phase transition becomes first order, similar to behavior seen in Ref.~\cite{coopparamagnet}.  This is likely due to vortices where the phase of $\phi$ winds by $\pi$, which are  more relevant at long length scales than the usual $2\pi$ vortices.
Close to the critical point, the transition should return to being second order at non-zero temperature.  However, the presence of low energy $Z_2$ magnetic vortices has a strong effect on the non-zero temperature properties.

The scenario that is consistent with the numerical data is that
for $V\gg t$, the electric defects have an energy of order $V$ and so are of much higher energy than the magnetic defects.  Increasing $t$  reduces the energy of the electric defects due to gain in kinetic energy, until this energy vanishes at the transition.  However, even a few magnetic defects can strongly influence the winding number distribution; the number of magnetic defects is roughly $L^2 \exp(-{\rm const.}/T)$, and so for $T\sim 1/L$, the number of magnetic defects is exponentially suppressed, but for moderate $L$, the prefactor leads to a substantial effect manifest as slow convergence of the scaling function at small $\beta/L$ as shaded in orange in Fig.~\ref{fig:wind}.

{\bf Discussion.} We have conclusively established the existence of a phase transition with deconfined fractionalized excitations in quantum Monte Carlo simulations of a physical model of lattice Bosons.  The topological properties of this transition are demonstrated by universal scaling functions of the winding number.   This test of the fractionalized charge uses the full winding distribution, and may be regarded as a strongly interacting analogue of current fluctuation measurements used experimentally to test fractionalized charge, as in Ref.~\cite{Heiblum}.  The full winding distribution is in principle experimentally accessible, as it is the Fourier transform of the free energy as a function of flux through the torus.

One can imagine more general deconfined quantum critical points with other gapped gauge fields, such as a $Z_k$ gauge field.   Lattice models for such theories are lacking, but if found, the winding number distribution that we have introduced will present a clear test of the universality class.
The fact that topological properties remain important even at the
critical point suggests that such systems, despite being gapless,
might be useful for quantum computing applications in the future.

\bibliography{XYrefs}

\bibliographystyle{Science}

{\it Acknowledgments} The authors thank T. Senthil for enlightening discussions.  MBH thanks the Aspen Center for Physics for hospitality.
This work has been supported by the Natural Sciences and Engineering Research Council of Canada (NSERC) and the Swiss HP$^2$C initiative. 
Simulations were performed on the Brutus cluster at ETH Zurich and the computing facilities of SHARCNET.

\section{Supporting Online Material}

\subsection{XY Model Simulations} \label{SOM1}

\begin{figure}[ht]
\centerline{\includegraphics[width=0.55\columnwidth]{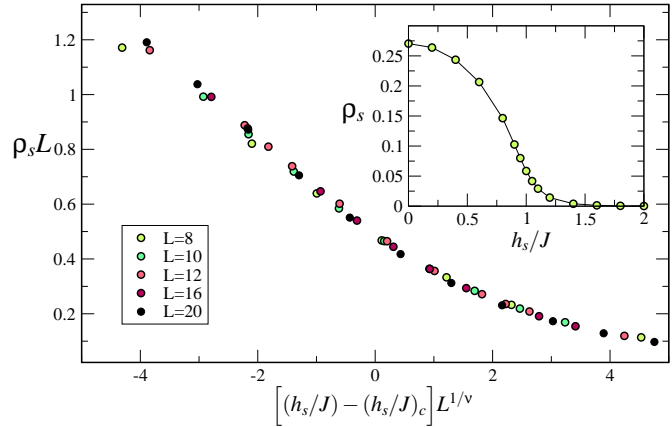}}
\caption{ 
Finite size scaling collapse of the superfluid density of the XY model in a staggered magnetic field, using $\nu = 0.6717$ and $(h_s/J)_c = 0.995$.
}
\label{fig:XYstagg}
\end{figure}

In this section, we describe quantum Monte Carlo (QMC) simulations of the two-dimensional XY model with an applied staggered field:
\begin{equation}
H = -J \sum_{\langle i j \rangle} (S^x_i S^y_j + S^y_i S^x_j) - h_s \sum_i  S^z_i \left({-1 }\right)^{x_i + y_i} 
\end{equation}
where ${\bf S_i}$ is a spin-1/2 operator, and $x_i$ and $y_i$ are the lattice coordinates of each spin.  
This Hamiltonian was chosen as a simple quantum lattice model to realize a transition in the $XY$ universality class between superfluid and insulating phases.  It has $XY$ symmetry but no higher continuous symmetry (in contrast, the $XXZ$ model has $SU(2)$ symmetry at the Heisenberg point), and applying a staggered field rather than a uniform field was chosen to avoid producing a chemical potential for bosonic excitations of the model which would produce terms linear in the time derivative for the effective action of the bosonic field.
We employ the stochastic series expansion QMC method with directed loops (the code was checked on a $4 \times 4$ lattice against Lanczos diagonalization), to study the phase transition out of the superfluid phase with increasing $h_s/J$.  The phase transition is identified through finite-size scaling of the superfluid density, $\rho_s$, as illustrated in Figure~\ref{fig:XYstagg}.  The broad features of the superfluid density identify a transition near $h_s \approx 1$ (inset), which through a finite-size scaling collapse is seen to be consistent with a continuous quantum phase transition with exponents $z=1$ and $\nu = 0.6717$ -- the values for the 3D XY universality class.  With the systems sizes studied, the best collapse of the finite-size data occurs at a critical $h_s/J = 0.995(5)$.

\begin{figure}[ht]
\centerline{\includegraphics[width=0.92\columnwidth]{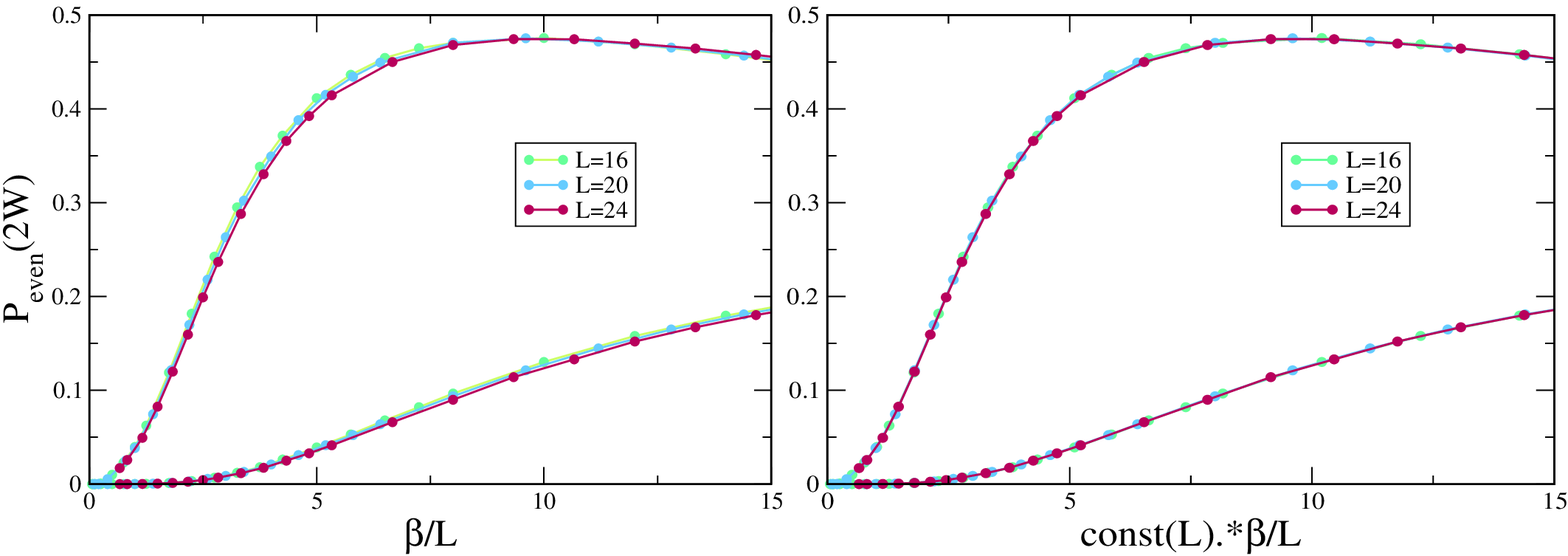}}
\caption{ 
Left: Plot of projected winding numbers, $P_{even}(2W)$ as a function of $\beta/L$ for $L=16,20,24$.  Right: Plot of projected winding numbers, $P_{even}(2W)$ as a function of $\beta$ for $L=16,20,24$, with small non-universal scaling applied to $x$-axis for each size.  Upper curves are $W=1$ and lower curves are $W=2$.
}
\label{fig:XYwind}
\end{figure}

Simulations were run to extract the probability distribution of the winding number squared at $h_s/J = 0.995$.  Measurements of both the full distribution were considered, as well as the ``projected'' distribution.  In the latter case, data was recorded only when both ($x$ and $y$) winding number were even (or zero), and the particle number was even.  These distributions allow us to calculate both the superfluid density of the physical bosons, as well as the ``projected'' space of composite bosons, which simulates the composite bosons of the fractionalized model (Fig.~\ref{fig:XYwind}).

We have tested the universality of the winding number distribution, to verify the hypothesis that at the critical point this converges to a universal function of
$v\beta/L$.  In Fig.~\ref{fig:XYwind}, we plot a scaling collapse of the projected winding distribution functions $P_{even}(2W)$ as a function of $\beta/L$ for $L=16,20,24$.
Only very minor differences can be seen between the collapsed curves; as a further test we plot the same winding distributions but allow additional non-universal factors (of the order of $2\%$) to multiply the $x$-axis for each system size.  As can be seen in  Fig.~\ref{fig:XYwind}, the collapse can be made almost perfect with these factors, indicating that the dominant finite size effect is a slight $L$-dependent correction to the velocity.  Since our comparison to the kagome lattice model involves multiplying by a non-universal velocity ratio, such $L$-dependent corrections to the velocity are unimportant in comparing the projected $XY$ and kagome curves.

\subsection{ Low Energy Magnetic Excitations} \label{SOM2}
The theory of the $XY^*$ critical point predicts that the electrical particles, carrying $\pm 1/2$ charge, become gapless as one approaches the critical point from the $Z_2$ spin liquid.  Indeed, in order to have the correlation function $\langle b^\dagger(0) b(r) \rangle$ not decay exponentially, some excitation carrying charge must become gapless.
The theory also predicts that magnetic particles remain gapped as the critical point is approached.  Once the electric particles condense in the superfluid phase, the magnetic excitations become confined and vanish from the spectrum of the theory.

At a non-zero temperature, however, there is some density of thermally excited magnetic particles in the spin liquid phase and at the critical point.
In this subsection we consider the properties of these excitations.  In particular, one complication is that while their excitation gap is predicted to remain non-zero at the transition, the actual numerical value is quite small, meaning that they can have a significant effect even at what seems to be a relatively low temperature.

\begin{figure}[ht]
\centerline{\includegraphics[width=0.5\columnwidth]{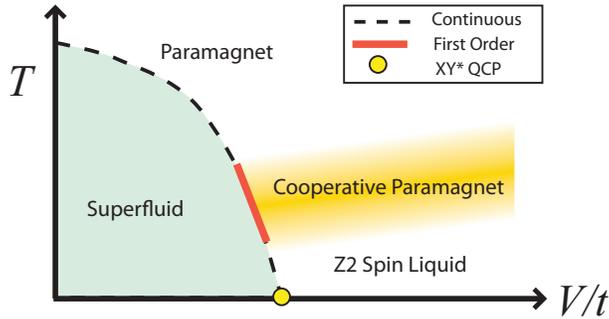}}
\caption{ 
Schematic phase diagram of the kagome Bose-Hubbard model. 
}
\label{fig:phase:diagram}
\end{figure}

First, consider the $Z_2$ spin liquid.  The study of \cite{IHM} showed two plateaus in the topological entanglement entropy as a function of temperature.  These plateaus are characteristic of two different energy scales, one for electric and one for magnetic excitations\cite{chamon,jiannis}.
Deep in the $Z_2$ spin liquid side with $t \ll V$, clearly the lower energy scale corresponds to magnetic excitations, that energy scale being set by $t^2/V$, while
the electric excitation energy scale is set by $V$.  However, as $t/V$ is varied, the evidence suggests that the characteristic energy scale of the magnetic excitations remains quite small.  This implies that at fairly small temperature there still is a non-zero density of magnetic excitations.  However, close to the critical point, we are confronted with an apparent paradox: the upper plateau appears to be at a fairly high temperature, and appears to correspond to the electric excitations.  One piece of evidence for this is based on the fact that the fluctuations in particle number are small (and correspondingly the uniform susceptibility described below is small) at this temperature scale, and also the fact that the density of defect hexagons becomes small at this scale.
However, the energy for the electric excitations must go to zero as they are the excitations that carry non-zero charge.
The resolution of this paradox that we propose consistent with the data is that the energy of the electric excitations does go to zero, but that at a small non-zero temperature there is also a density of magnetic excitations.  These magnetic excitations renormalize the energy of the electric excitations upwards (the energy of an electric excitation can be thought of as a balance between a positive term due to creating a defect hexagon and a negative term due to gain in kinetic energy, but the magnetic excitations reduce the kinetic energy gain), and so we see a regime at intermediate temperature which has a large density of magnetic excitations but a low density of electric excitations.  This regime  is similar to the cooperative paramagnet phase\cite{coopparamagnet}; it is not separated from the high temperature paramagnet by a phase transition but there is a very distinct, measurable crossover in properties such as the topological entanglement entropy and the susceptibility.

\begin{figure}
\centerline{\includegraphics[width=0.55\columnwidth]{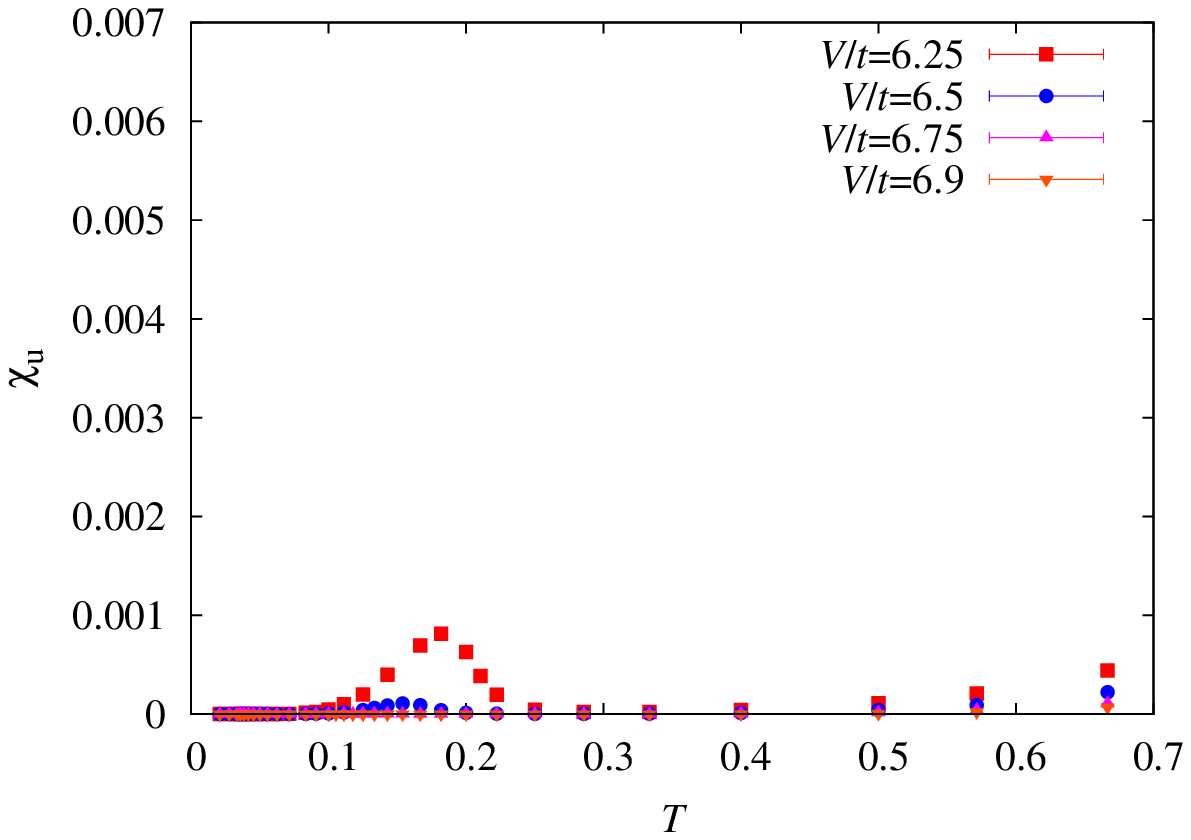}}
\centerline{\includegraphics[width=0.55\columnwidth]{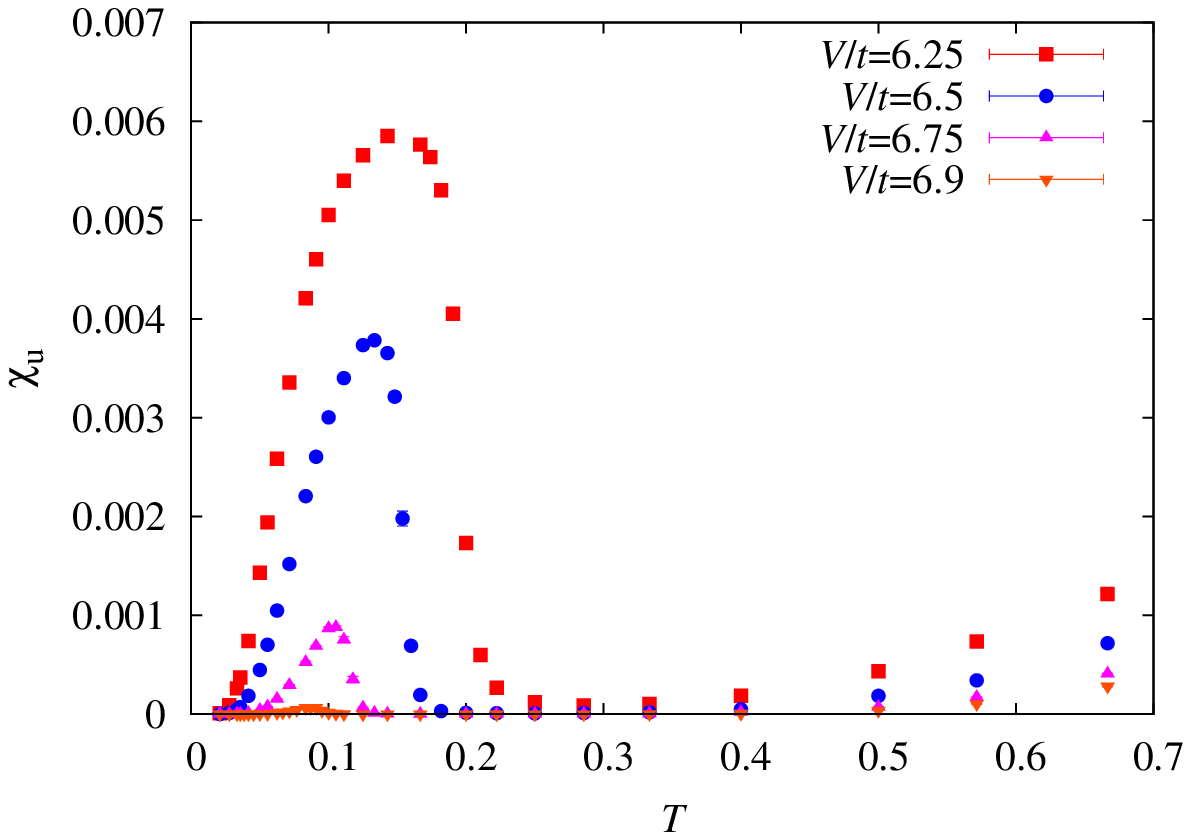}}
\centerline{\includegraphics[width=0.55\columnwidth]{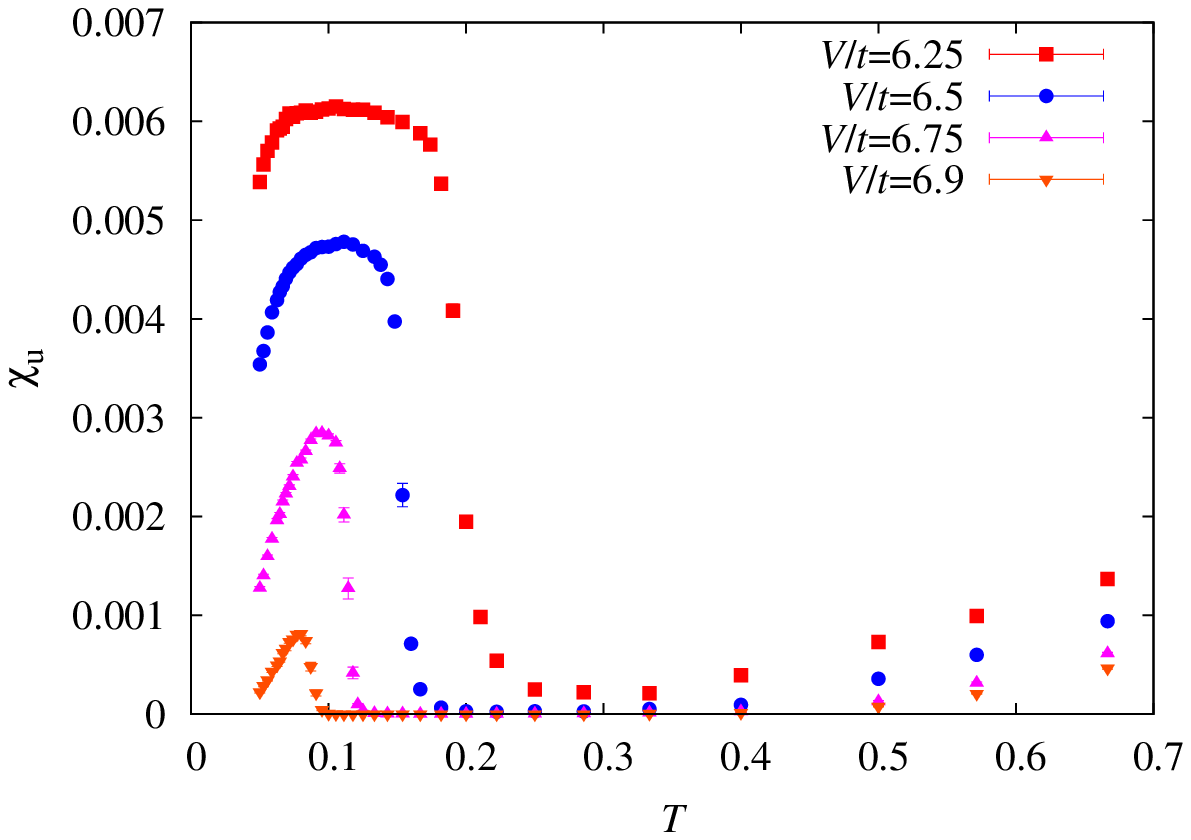}}
\caption{ 
The uniform susceptibility $\chi_u$ as a function of temperature for different values of $V/t$ on the superfluid side of transition. Upper panel: $L=8$. Middle panel: $L=16$. Lower panel: $L=24$.
}
\label{fig:chi}
\end{figure}

These magnetic excitations also have an important effect on the properties for $V/t<(V/t)_c$, in the superfluid phase.  Consider the phase diagram as a function of $V/t$ and temperature, Fig.~\ref{fig:phase:diagram}.  At small $V/t$ we see a continuous phase transition as a function of temperature.  This transition is likely Kosterlitz-Thouless, but closer to the critical point the transition appears to become first order (see also Ref.~\cite{coopparamagnet}).  As discussed in the text, this is likely due to the existence of vortices where the phase of the $\phi$ field winds by $\pi$.  However, such phase winding corresponds to binding a $Z_2$ vortex to the phase vortex and hence the importance of these vortices depends upon the energy scale of the magnetic excitations.
Since the $T=0$ phase transition is continuous as a function of $V/t$, eventually we expect that for $V/t$ sufficiently close to $(V/t)_c$ the transition as a function of $T$ will become continuous again, but we have not fully resolved this region.

\begin{figure}
\centerline{\includegraphics[width=0.55\columnwidth]{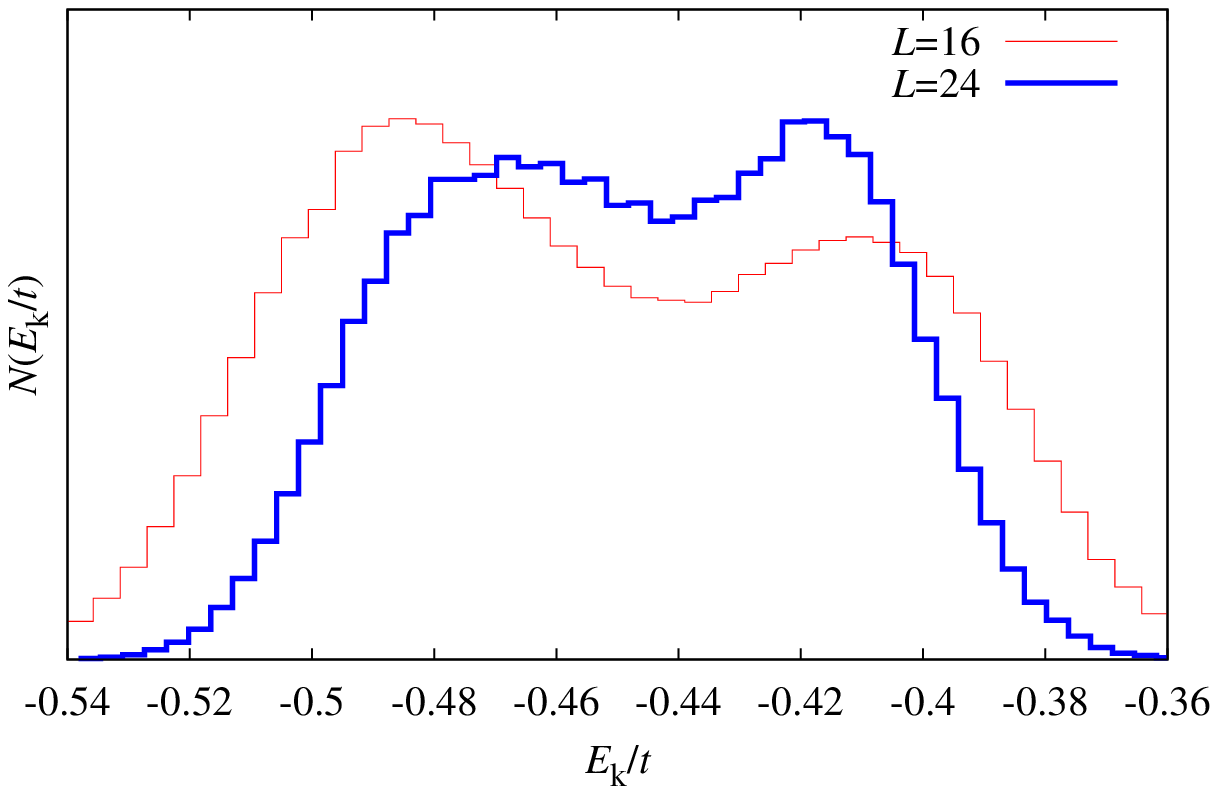}}
\caption{ 
Distribution of the kinetic energy per site close to the finite-temperature transition at $V/t=6.5$ and $\beta=7.5$. 
}
\label{fig:histogram}
\end{figure}

Evidence for the existence of the first order transition can be obtained, for example, by studying the uniform susceptibility as a function of $V/t$ and temperature.  This susceptibility is obtained from fluctuations in the particle number by
\be
\chi_{u}\equiv \frac{1}{L^2 T} \langle (n-\overline n)^2 \rangle,
\ee
where $n$ is the total boson occupation number. As shown in Fig.~\ref{fig:chi}, we see a rise in susceptibility at low temperature, followed by a sudden decline.  We associate this with a first order transition to a cooperative paramagnet. More direct evidence is a visible double peaked structure of the distribution of the kinetic energy, $E_k=-t\langle\sum_{ij} (b_i^\dagger b_j+b_i b_j^\dagger)\rangle$ -- see Fig.~\ref{fig:histogram}.

\begin{figure}
\centerline{\includegraphics[width=0.55\columnwidth]{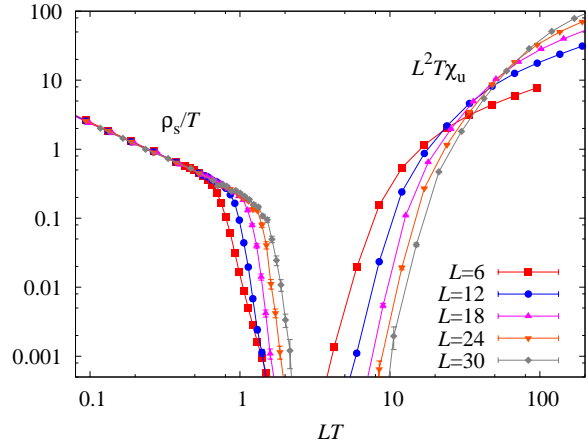}}
\caption{ 
Superfluid density $\rho_s$ and the uniform susceptibility $\chi_u$ as a function of $LT$.  The uniform susceptibility is strongly suppressed at low and intermediate temperatures.  Lines guide the eye.
}
\label{fig:ltscaling}
\end{figure}

The existence of these magnetic excitations poses one further issue for analyzing the critical point.  One might hope to follow a similar approach to Ref.~\cite{KaulMelko}, where the uniform susceptibility and superfluid density are both considered as function of $\beta/L$, and universal amplitudes are extracted from considering the combination of these two quantities.  The superfluid density becomes non-negligible once $\beta/L$ is large, while the uniform susceptibility is non-negligible when $\beta/L$ is small (since the gap to a particle excitation is of order $1/L$, the particle number variance is exponentially small in $\beta/L$).  If the limit of large $L$ is taken at fixed $\beta/L$, then we also take the large $\beta$ limit and hence we will eventually reach a temperature scale low enough to have a negligible number of magnetic excitations. Note that the total number of magnetic excitations is roughly $L^2 \exp(-\beta \Delta E)$, where $\Delta E$ is the energy gap to magnetic excitations, and hence this number decays exponentially in $L$ assuming fixed $\beta/L$.  However, in the regime where the uniform susceptibility is non-negligible, which corresponds to a small value of $\beta/L$, we find that there still is a non-negligible number of magnetic excitations, due to the $L^2$ prefactor, at accessible values of $L$.  For this reason, as shown in Fig.~\ref{fig:ltscaling}, we do not see a good scaling collapse at small and intermediate $\beta/L$.  One may expect that a scaling collapse might be recovered in some range at large values of $\beta/L$ for large enough system sizes $L$, when the gap to electric excitations becomes well bellow the gap to magnetic excitations. Some evidence is shown in Fig.~\ref{fig:chi}, where one can see that the uniform susceptibility is strongly suppressed for small system sizes but it becomes finite for large enough system sizes as the two energy scales cross over.  The effect gets more pronounced as we approach the quantum critical point.

While this means that we cannot find universal properties by analyzing $\chi_{\mu}$ and $\rho_s$ in parallel, the winding number distribution measurement discussed in the main text provides a way of extracting the universal properties of the fractional charge by studying properties that are determined by properties at a sufficiently low temperature to remove the magnetic excitations.

\end{document}